\documentclass[prb  ,aps,showpacs,twocolumn,unsortedaddress,longbibliography,superscriptaddress]{revtex4-1}
\usepackage{latexsym}
\usepackage{graphicx}
\usepackage{color}
\usepackage{amsmath}
\usepackage{soul}

\setcitestyle{super,open={},close={}}

\makeatletter
\renewcommand\@biblabel[1]{#1.}
\makeatother

\begin{document}
\title{Buried Dirac points in quantum spin Hall insulators: Implications for Majorana Kramers pair-based quantum computing}

\author{J.J.~Cuozzo}\email{jjcuozz@sandia.gov}\affiliation{Department of Physics, The University of Texas at El Paso, El Paso, TX, USA.}\affiliation{Sandia National Laboratories, Livermore, CA 94551, USA.}
\author{W.~Yu}\affiliation{Sandia National Laboratories, Albuquerque, New Mexico 87185, USA}
\author{X.~Shi}\affiliation{Department of Physics, The University of Texas at Dallas, Richardson, TX, USA}
\author{A.J.~Muhowski}\affiliation{Sandia National Laboratories, Albuquerque, New Mexico 87185, USA}
\author{S.D.~Hawkins}\affiliation{Sandia National Laboratories, Albuquerque, New Mexico 87185, USA}
\author{J.F.~Klem}\affiliation{Sandia National Laboratories, Albuquerque, New Mexico 87185, USA}
\author{E.~Rossi}\affiliation{Department of Physics, William \& Mary, Williamsburg, VA 23187, USA}
\author{W.~Pan}\email{wpan@sandia.gov}\affiliation{Sandia National Laboratories, Livermore, CA 94551, USA.}

\date{\today}
\begin{abstract}
Quantum spin Hall insulators (QSHI) host helical electronic edge states that are protected from backscattering due to time-reversal symmetry (TRS).
Despite considerable work investigating QSHI edge states, there is still an open question about their unexpected resilience to large magnetic fields where TRS is undoubtedly broken.
In this work, we investigate the transport properties of helical edge states in a QSHI-SC junction formed by a InAs/GaSb (15nm/5nm) double quantum well and a superconducting tantalum (Ta) constriction.
We observe a robust conductance plateau up to 2~T, signaling resilient edge state transport. 
Using a modified Landauer-B\"{u}ttiker analysis, we find that the zero field conductance is consistent with 98\% Andreev reflection probability owing to the high transparency of the InAs/GaSb--Ta interface. 
Such resilience is consistent with the Dirac point for the edge states being buried in the bulk valence band.
We further theoretically show that a buried Dirac point does not affect the robustness of the quasi-1D topological superconducting phase.
We find a buried Dirac point favors the hybridization of Majorana Kramers pairs (MKPs)-- predicted to exist in a QSHI-SC constriction-- and fermionic modes in the QSHI vacuum edge resulting in \textit{extended} MKP states, highlighting the subtle role of buried Dirac points in probing MKPs.
\end{abstract}

\maketitle

\section{Introduction}
\begin{figure*}[t]
    \centering
    \includegraphics[width=0.8\linewidth]{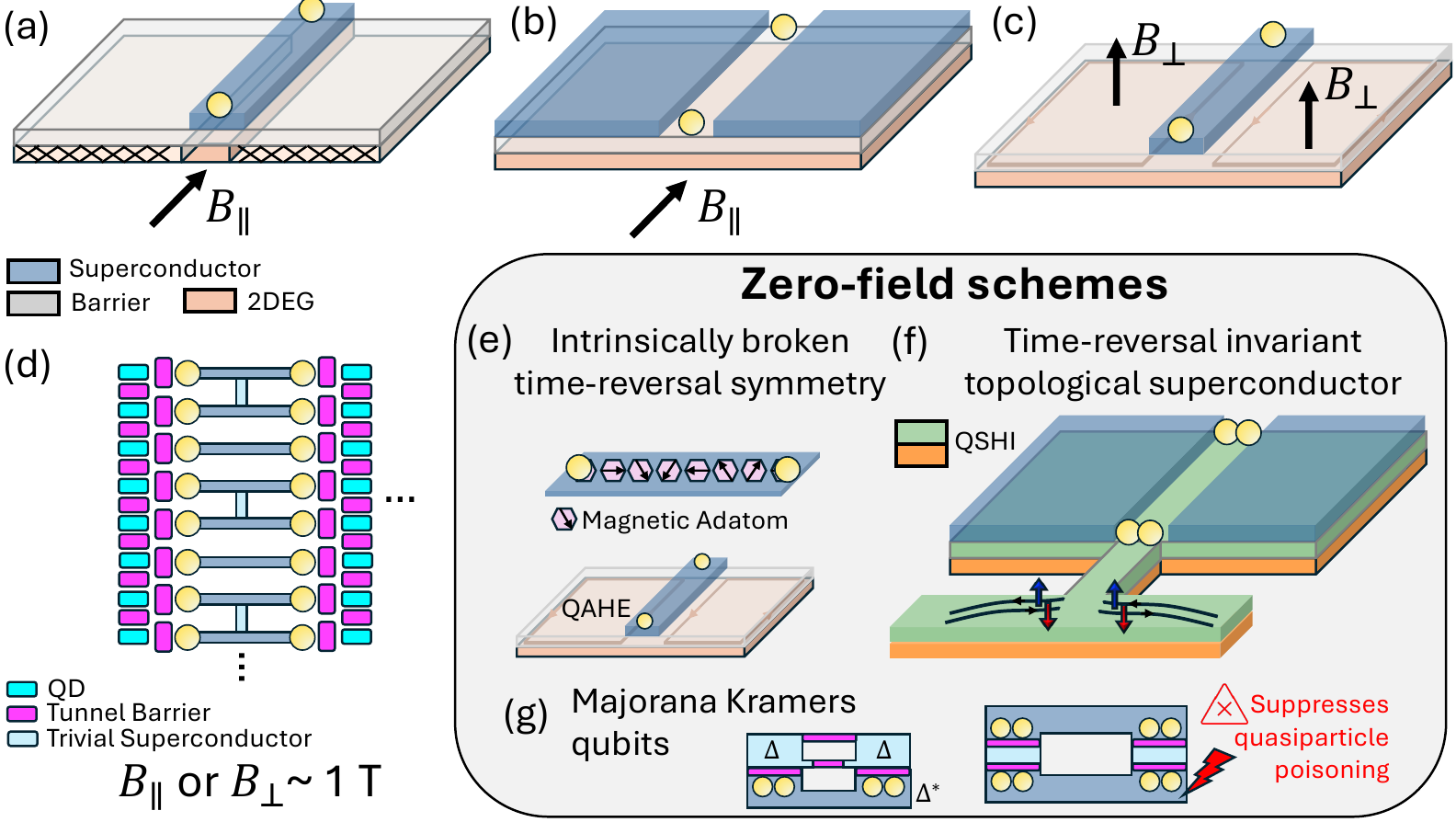} 
    \caption{\label{fig:fig0}
    Schematic of TSCs with MZMs (yellow dots) in \textbf{(a)} a gate-defined Lutchyn-Oreg nanowire and
    \textbf{(b)} planar Josephson junction with applied in-plane magnetic field $B_{\parallel}$, and
    \textbf{(c)} quantum Hall-superconductor hybrid structure with out-of-plane field $B_{\perp}$.
    \textbf{(d)} Tetron qubit-based topological quantum computing architecture based on arrays of coupled TSCs.
    Zero field schemes for TSC:
    \textbf{(e)} Schematic of approaches relying on intrinsically broken TRS with magnetic adatoms on the surface of a conventional superconductor forming a Shiba chain, or quantum anomalous Hall effects that do not require an external magnetic field.
    \textbf{(f)} Time-reversal invariant TSC based on a QSHI and SC constriction which hosts two degenerate pairs of Majoranas (Majorana Kramers pairs). 
    \textbf{(g)} Schematic of a Majorana Kramers qubit based on Majorana Kramers pairs than can perform (left) single- and (right) two-qubit gate operations required for universal quantum computing.
    }
\end{figure*}

Quantum materials are currently under intense investigation to understand
fundamental interactions driving quantum phases of matter
which can be leveraged for Quantum Information Science and to create novel quantum technology~\cite{Cava2021, Schiela2024}.
Quantum materials are distinct from ``classical" materials in that they exhibit emergent quantum phenomena with no classical analogs~\cite{Nature_quantum_mat} such as topological protection arising from the bulk bandstructure of a topological insulator (TI)~\cite{Hasan2010}.
In the seminal work by Kitaev, it was recognized that topological protection could be utilized to suppress quantum decoherence in a quantum computer, for example by performing qubit gate operations with topological defects in a so-called Kitaev chain-- a one-dimensional spinless p-wave superconductors belonging to class D of the Altland-Zirnbauer classification~\cite{Kitaev2001}. These exotic topological defects are called Majorana zero modes (MZM) and can be used to construct topological qubits satisfying requirements for universal quantum computing~\cite{ Kitaev2003, Nayak2008, Alicea2012}. 
Since the initial work of Kitaev, many proposals have been offered to realize a Kitaev chain and other analogs hosting MZMs;
however, their unambiguous experimental demonstration remains elusive~\cite{Lee2012_zb_anom, Kells2012, Reeg2018, Vuik2019, Liu2019, Chen2019, Awoga2019, Woods2019, Valentini2021, Hess2021, microsoft2022, microsoft2025}.
Most of the approaches to creating a topological superconductor (TSC) with MZMs rely upon engineering a heterostructure involving conventional superconductors and the superconducting proximity effect. Within this set of proposals, the setups can be distinguished into two sub-groups: finite and zero magnetic field schemes. 

Three popular approaches to creating MZMs involving finite magnetic fields are described in Fig.~\ref{fig:fig0}~(a-c) and capture the essence of this sub-group of schemes. Figure~\ref{fig:fig0}~(a) describes the Lutchyn-Oreg nanowire model~\cite{Lutchyn2010, Oreg2010} that utilizes a semiconducting nanowire with strong spin-orbit interactions (e.g. InAs), a moderate axial magnetic field $B_{\parallel}$ ($B_{\parallel} < 1$~T), and a conventional superconductor (e.g. Al). A similar construction is shown Fig.~\ref{fig:fig0}~(b) involving a planar Josephson junction~\cite{Pientka2017, Hell2017}. Here the essential distinction from the Lutchyn-Oreg picture is how a phase-bias across the junction can be used to create a stable topological superconductor (TSC) inside the junction. Lastly, Fig.~\ref{fig:fig0}~(c) describes
the approach using the quantum Hall effect.
In this case, a strong magnetic field ($B_{\perp} > 1$~T) drives a two-dimensional electron gas (2DEG) in the quantum Hall regime, whose counter-propagating edge states can couple through a superconductor to create a TSC~\cite{Fu2008, Fu2009, Tanaka2009, Mong2014, SanJose2015, Lee2017, Zhao2020, Hatefipour2022, gul2021induced, Cuozzo2024_andreev, Hatefipour2024}. While these various approaches are viable paths to creating a TSC with MZMs, their integration into a quantum computing architecture~\cite{microsoft_roadmap2025} like the one shown in Fig.~\ref{fig:fig0}~(d) will inevitably involve many TSCs relying on a global magnetic field on the order of one Tesla, creating challenges for performance and scalability.

Alternatively, the zero-field schemes described in Fig.~\ref{fig:fig0}~(e-g) do not rely upon an applied magnetic field. Figure~\ref{fig:fig0}~(e) describes two approaches where spin textures are induced in a superconductor, e.g. a Shiba chain~\cite{Nadj-Perge2013, Nadj-Perge2014}, or where zero-field analogs of the approaches described in Fig.~\ref{fig:fig0}~(a-c) are implemented, e.g. using quantum anomalous Hall insulators~\cite{Clarke2014, Uday2024}. 
While an external magnetic field is not necessary to realize MZMs, time-reversal symmetry has to be broken in the approaches such as shown Fig.~{\ref{fig:fig0}}~(e), that, however, are also very difficult to scale requiring atomic precision fabrication.
In this work, we explore an approach that is fundamentally distinct where four-fold degenerate MZMs are realized in a time-reversal \textit{invariant} topological superconductor based on a quantum spin Hall insulator (QSHI), see Fig.~\ref{fig:fig0}~(f).

A QSHI is a 2D TI hosting a pair of time reversal symmetry-protected helical edge states~\cite{Hasan2010, Qi2011, Kane2005, BHZ2006}.
As shown in Fig.~\ref{fig:fig0}~(f), when helical edge states encounter a superconducting constriction with a $\pi$-phase bias, Majorana Kramers pairs~\cite{Teo2010} (MKPs) can form at the ends of the constriction~\cite{Li2016, Pikulin2016}.
Recently, it has been suggested that MKPs can be used to create a Majorana Kramers qubit (MKQ), see Fig.~\ref{fig:fig0}~(g), that can perform a set of universal single- and two-qubit logic gates~\cite{Schrade2022}. Importantly, MKQs may benefit from preserved time-reversal symmetry by having a larger topological gap and additional protection from quasiparticle poisoning, which could make them a viable alternative to MZM-based qubits~\cite{Schrade2022}.

Currently, there exist experimental challenges to realize a high quality QSHI-SC heterostructure, such as edge roughness causing spurious edge conduction~\cite{Nichele2016}. Furthermore, QSHI devices with a quantized Hall conductance have shown a confounding resilience to large magnetic fields~\cite{Du2015}, complicating the edge state picture where broken TRS is expected to open a gap at the Dirac point of linearly dispersing edge states, which ought to lead to the suppression of edge conductance. One hypothesis to resolve this discrepancy is to consider higher-order corrections to the band structure of QSHIs which can lead to a burying of the Dirac point in the bulk valence bands, i.e., well below the bulk mid-gap energy~\cite{Li2018,Skolasinski2018,Krishtopenko2018}. This is similar to buried Dirac points reported in 3D topological insulators like Bi$_2$Te$_3$~\cite{Zhang2009},
and these works highlight the need to implement models that go beyond simplified minimal models of QSHIs in zincblende systems.
Recent work theoretically investigated the impact of a buried Dirac point on a QSHI-SC heterostructure with a ferromagnetic layer~\cite{Schulz2020}, but a study of MKPs is currently lacking.

\begin{figure*}[t]
    \centering
    \includegraphics[width=0.9\linewidth]{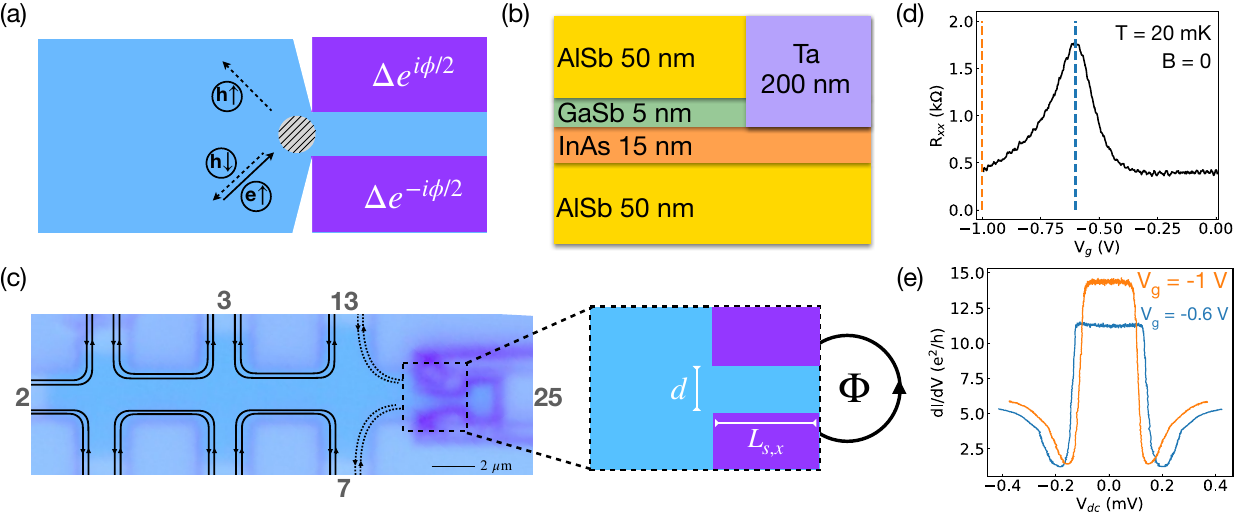} 
    \caption{\label{fig:fig1}
    \textbf{(a)} Schematic of Andreev retroreflection and crossed Andreev reflection at a QSHI-SC constriction.
    \textbf{(b)} Illustration of the material stack. 
    \textbf{(c)} Microscope image of the device with edge states schematically superimposed. Contact 25 is the Ta SC QPC and all other contacts are Ti/Au. Dashed box: Schematic of the superconducting point contact formed by the junction.
    \textbf{(d)} Longitudinal resistance R$_{xx}$ across normal contacts 3 and 13 vs gate voltage.
    \textbf{(e)} Three-terminal $dI/dV$ vs voltage bias between contacts 13 and 25 at $B=0$ and $V_g = -0.6,~-1.0$~V
    }
\end{figure*}

In this work we present a device deliberately designed toward the goal of realizing MKPs. This requires the combination of several features: (i) Extremely high quality InAs/GaSb semiconductor stacks grown via molecular beam epitaxy to minimize the effects of disorder, that has emerged as the main obstacle toward the realization of non-abelian states and therefore topological qubits; (ii)  accurate control of the thickness of the InAs and GaSb to be in the “inverted-band” regime in which the 2DEG is expected to be in a QSHI phase; (iii) nanoscale fabrication of a superconducting-2DEG hybrid structure and patterning of a superconducting constriction to allow the trapping of the MKPs without decreasing the quality of the 2DEG;  and (iv) achievement of almost ideal superconducting edge states at the constriction.
When incoming electrons along the QSHI edges meet a phase-biased SC constriction, the process of Andreev reflection can occur depending on the phase difference across the constriction, see Fig~\ref{fig:fig1}~(a).
To probe these Andreev processes, we measure a quantized three-terminal conductance of $12(e^2/h)$ in an InAs/GaSb(15nm/5nm) double quantum well across a superconducting Ta constriction. 
Analyzing the conductance with a modified Landauer-B\"uttiker model accounting for Andreev reflection of helical edge states, we find the conductance measured is consistent with nearly perfect Andreev retroreflection ($98\%$), owing to the high quality of the device and the helical edge state properties. 
Performing numerical simulations of edge state transport using the Bernevig-Hughes-Zhang (BHZ) model~\cite{BHZ2006} within the Bogoliubov de-Gennes (BdG) formalism, we find quantitative agreement with measured conductance at zero field.
In addition, we present results that show that the Dirac point of the edge states in the InAs/GaSb double quantum well is buried in the bulk conduction band, and that this makes the superconducting edge modes needed to realize MKPs robust against external magnetic fields. We then show theoretically that the fact that the Dirac point is below the edge of the bulk conduction band does not affect the topological gap protecting the MKPs. Consequentially, we establish the device design presented as a viable, concrete, path toward the realization of MKP and time-reversal invariant topological quantum computing.

\section{Device fabrication}
In this study the QSHI is realized in InAs/GaSb double quantum wells. InAs and GaSb have a small effective mass, high mobility, and low Schottky barrier to superconductors like Ta, making them attractive for superconducting heterostructure device design. 
Molecular beam epitaxy techniques are used to grow a InAs/GaSb (15nm/5nm) double quantum well. This thickness arrangement of the double quantum well corresponds to the inverted regime where a QSHI develops~\cite{Yu2018_exciton}. The growth recipe is similar to what has been done before~\cite{Yang1996} and has been described in Refs. \citenum{Yu2014} and \citenum{Shi2015}. A cross-sectional cut of the stack and the Ta electrode is shown in Fig.~\ref{fig:fig1}~(b) (full stack shown in Appendix~\ref{appendix:exp}).
The AlSb layers act as a potential barrier for the InAs/GaSb bilayer.
We define the InAs/GaSb mesa using standard photolithography techniques and wet chemical etching~\cite{Yu2014}.
We use the etchant NH$_4$OH (22\%) : H$_2$O = 1: 4 to  etch GaSb and AlSb (etching rate ~ 4 \AA/s), and Citric Acid : H$_2$O$_2$ = 1:2 to etch InAs (etching rate ~ 4 \AA/s). These etchants are quite selective, so we normally use longer etching time to make sure that each layer is completely etched off.
Then a Ta electrode is directly deposited in the pattern of an rf-SQUID, as shown in Fig.~\ref{fig:fig1}~(c), creating a superconducting point contact (SPC) in the InAs/GaSb mesa.
To complete device fabrication, an Al$_2$O$_3$ dielectric layer is grown by atomic layer deposition, before depositing a metal stack of Au/Ti (300nm/20nm). Biasing the top Au/Ti gate allows tunability of the electron density in the bilayer. 
Helical edge states propagating along the InAs/GaSb mesa encounter the constriction formed by the Ta junction and hybridize, opening up a mass gap in the edge dispersion. 
In this device, the constriction length is $L_{s,x} \approx 2~\mu$m and the separation of the Ta arms is $d =$ 600 nm (Fig.~\ref{fig:fig1}~(c)). 
The edge states of InAs/GaSb
have been shown to have a spatial penetration into the bulk with an upper bound of about 300 nm~\cite{Pribiag2015}, so that the constriction of the device ($d \approx 600$ nm) should lead to significant scattering between edges.
The penetration of the edge states in the bulk roughly scales inversely with the topological bulk gap. With our numerical simulations of InAs/GaSb presented in Sec.~\ref{sec:mkp}, we find an edge penetration of about 300 nm. Since the bulk gap in the simulations are consistent with the bulk gap in 15/5 nm InAs/GaSb, we suppose our 300 nm upper bound is reasonable.

\section{Experimental Data}
The carrier density we obtained at $V_g = 0$~V (which is close to the as-grown density) is $3.85 \times 10^{11}$~cm$^{-2}$. This as-grown density is similar to those reported in previous works in similar growth structures~\cite{Pan2013, Nichele2014}.
The measurements reported in this work were done on one device where we have verified that we get the same results on different cool downs.
In Fig.~\ref{fig:fig1}~(d), we show the measured longitudinal resistance $R_{xx}$ as a function of the electrostatic gate voltage $V_g$ at base temperature $T = 20$~mK. A peak occurs around $V_g \approx -0.6$ V, which we attribute to the presence of the bulk gap in the InAs/GaSb bilayer. The maximum resistance is $\sim 1.75$ k$\Omega$, which is roughly 13\% of the resistance quantum $R_0 = h/2e^2$. Based on a resistor network model~\cite{Mueller2015}, we expect a resistance of $R_{xx} = 3R_0/8 \approx 4.8$~k$\Omega$ in the QSHI phase. We believe this discrepancy is primarily due to a non-ideal contact from the Ta electrode; although, there is likely also some bulk conductivity associated with disorder in the InAs/GaSb bilayer, see Appendix~\ref{appendix:exp}. 

\begin{figure}[t]
    \centering
    \includegraphics[width=0.98\linewidth]{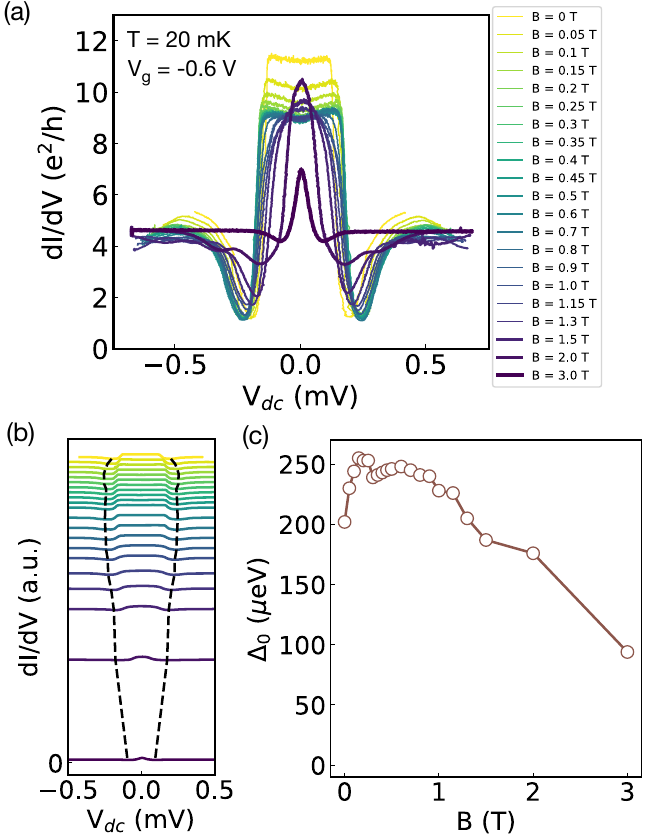} 
    \caption{\label{fig:fig2}
    \textbf{(a)} Three-terminal $dI/dV$ vs $V_{dc}$ under an applied magnetic field $B$.
    \textbf{(b)} Waterfall plot of $dI/dV$ with increasing $B$. Dashed line follows the evolution of the superconducting gap $\Delta_0$.
    \textbf{(c)} Gap $\Delta_0$ extracted from the $dI/dV$ traces vs magnetic field.
    }
\end{figure}

To investigate the edge state regime further, in Fig.~\ref{fig:fig1}~(e) we compare three-terminal conductance measurements across the Ta constriction at the gate voltage corresponding to the peak in R$_{xx}$ ($V_g = -0.6$~V) and in the hole-doped regime ($V_g = -1.0$~V). Here a current bias is applied between electrodes 7 and 25, and voltage is measured between 25 and 13. Both regimes show a nearly constant conductance before the onset of conductance dips at $V_{dc} \approx \pm 0.2$ mV. 
Figure \ref{fig:fig2}~(a) shows the differential conductance at $V_g = -0.6$~V for various out-of-plane magnetic fields. It is clear that edge state transport appears to be robust to magnetic fields up to $\sim 2$~T.
Indeed, at 2~T, the zero-bias conductance is still robust despite a Zeeman splitting $\Delta_Z = g\mu_B B \approx 0.88$~meV (assuming a bulk g-factor of 8) which is almost half of the theoretical bulk gap of InAs/GaSb at $B=0$.  

Dips in the differential conductance at $V_{dc} \approx \pm 0.2$~mV are consistent with the superconducting gap of Ta ($\sim200$~$\mu$eV).
In Fig.~\ref{fig:fig2}~(b) we present the dependence of the gap extracted from the minima of $dI/dV$, $\Delta_0$, as a function of $B$. 
Firstly, we observe a slight enhancement of the gap at low fields, similar to past measurements~\cite{Shi2015}. This may be associated with the strong spin-orbit interaction~\cite{Gardner2011} in the InAs/GaSb bilayer.
Notably, $\Delta_0$ is nearly constant at low fields and is gradually suppressed by the magnetic field.

\begin{figure*}[t]
    \centering
    \includegraphics[width=0.9\linewidth]{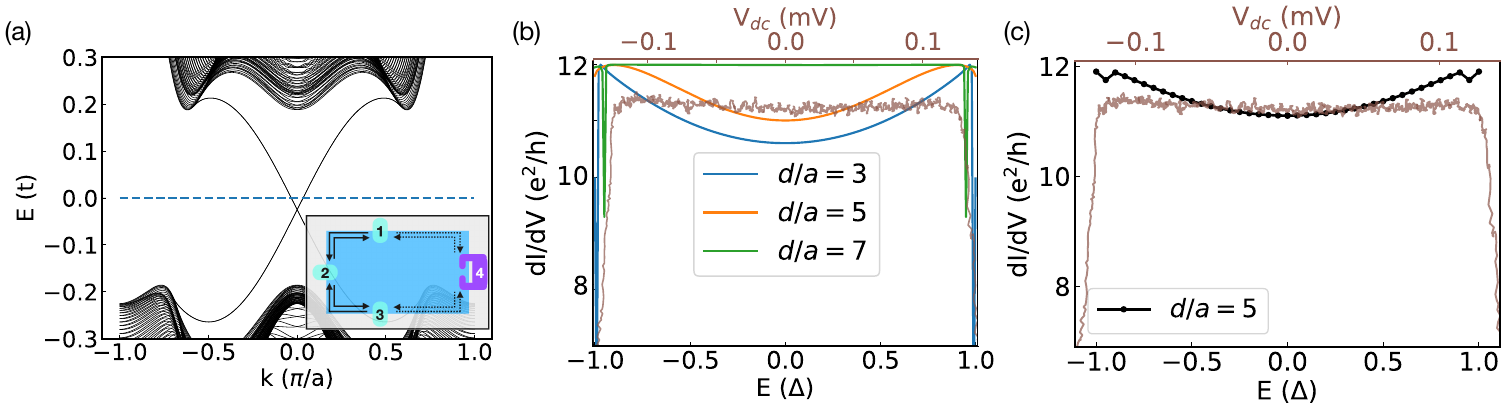} 
    \caption{\label{fig:fig3}
    \textbf{(a)} Dispersion of BHZ model used in transport simulations. Dashed line indicates the chemical potential used in simulations presented in panels (b,c).
    \textbf{(b)} $dI/dV$ comparison between experimental measurements (brown) and simulation with various constriction gaps $d$ for $L_y=120a$ and $L_{x,s}=100a$ where $a$ is the tight binding lattice spacing.
    \textbf{(c)} $dI/dV$ comparison between experimental measurements (brown) and simulation with Anderson disorder (black), $U_A = 60\Delta$ and $d=5a$.
    }
\end{figure*}

\section{Numerical Conductance Simulations}
To interpret our measurements, we simulate the electronic transport numerically with a three-terminal setup.
Within the bulk gap of a QSHI, nominally we expect a pair of counter-propagating helical edge states to be responsible for charge transport. When one of these edge states encounters a superconducting interface with an excitation energy below the superconducting gap, the process of Andreev reflection occurs where a hole is retroreflected from the superconductor and a pair of electrons co-tunnels into the condensate of Cooper pairs in the superconductor. 
Perfect Andreev reflection is expected since single electron tunneling into the superconductor and backscattering along the QSHI vacuum edge are forbidden.
In our device with a superconducting constriction, however, the geometry of the constriction can strongly affect the probability of Andreev reflection since backscattering and electron tunneling across the constriction edge are allowed.
We model this by calculating the scattering matrix coefficients using the python package Kwant~\cite{Groth2014}. First, we treat the superconducting constriction as two parallel superconducting electrodes separated by a small gap (Fig.~\ref{fig:fig1}~(a)). We can treat the QSHI-SC system within the BdG formalism as 
\begin{align}
    H_{BdG}(\mathbf{k}) &= 
    \begin{pmatrix}
        H_{bhz}(\mathbf{k}) - \mu && \hat{\Delta} \\
        \hat{\Delta}^* && \mu - T H_{bhz}(-\mathbf{k}) T^{-1}
    \end{pmatrix},
    \label{eq:Ham}
\end{align}
which satisfies the BdG equation 
$H_{BdG} \begin{pmatrix}
    u \\ v
\end{pmatrix} = E \begin{pmatrix}
    u \\ v
\end{pmatrix}$ 
and where the mean-field superconducting gap $\hat{\Delta}(\mathbf{r}) = \Delta e^{i\phi/2}$ in the bottom superconducting QPC arm and $\Delta e^{-i\phi/2}$ in the top arm, $\mu$ is the chemical potential, and 
$\mathbf{k} = -i\nabla_{\bf r}$ is the momentum operator in real space.
%
$T=i\sigma_y K$ is the time-reversal operator with complex conjugate operator $K$. The BHZ model describing the QSHI is 
$
H_{bhz}(\mathbf{k}) = H_{bhz,0} + \frac{\Delta_Z}{2} \sigma_z + H_{IA},
$
where
$
H_{bhz,0}(\mathbf{k}) = C + M \rho_z - (B \rho_z + D )k^2 + A (k_x \sigma_z \rho_x - k_y \rho_y)
$ is the conventional BHZ model,
$\Delta_Z$ is the Zeeman spin splitting, $H_{IA}$ describes broken inversion symmetry in the double quantum well~\cite{Liu2008}, $\rho_i$, $\sigma_i$ are Pauli matrices in orbital and spin space, respectively. A detailed description of the system is presented in Appendix~\ref{appendix:bhz}. The dispersion of the BHZ model is shown in Fig.~\ref{fig:fig3}~(a) where the linearly crossing bands at $E=0$ are the helical edge states.

To make the connection to our measured conductance in Fig.~\ref{fig:fig2}~(a), we calculate the sub-gap conductance of the four-terminal Hall bar shown in the inset of Fig.~\ref{fig:fig3}~(a) using a modified Landauer-B\"{u}ttiker model for helical edge modes. 
If we consider the possibility of spin-flip processes, then both backscattering of edge states and crossed Andreev reflection may occur in the device, so we include all of these processes in the model. We assume leads 2 and 3 are floating and lead 4 is a grounded superconductor. The Landauer-B\"{u}ttiker equation for the current in lead $i$ is
$
I_i = \sum_{j} a_{ij}(V_j - V)
$
where $V_i$ is the chemical potential of lead $i$, $V = V_4$, and
$
a_{ij} = R_0^{-1} \left( 2 - T^{ee}_{ij} + T^{he}_{ij} \right)
$
with the resistance quantum $R_0 = h/e^2$. Here, $T^{ee}_{ij}$ and $T_{ij}^{he}$ are probabilities of electron transmission and Andreev reflection from lead $j$ to lead $i$, respectively, and $0 \le T_{ij}^{ee},T_{ij}^{he} \le 1$. We suppose the currents $I_1 = I = -I_4$.
We can calculate the resistance $R_{34} = \frac{V_3 - V}{I}$ and find
\begin{align}
    R_{34} = R_0 \left[ \frac{1}{4 (T^{he}+T^{car})} - \frac{1}{2(3 - 2T^{car} - 2T^{b})} \right] 
    \label{eq:R34}
\end{align}
where $T^{he} = T^{he}_{11} = T^{he}_{33}$ is the Andreev reflection probability, $T^{car} = T^{he}_{13} = T^{he}_{31}$ is the crossed Andreev reflection probability, and $T^{b} = T^{ee}_{11} = T^{ee}_{33}$ is the backscattering probability. The conservation of quasiparticle current is expressed as $T^{ee}_{11}+T^{he}_{31} + T^{he}_{11}+ T^{ee}_{31} = 1$. Details of this calculation are provided in Appendix~\ref{appendix:transport}.
At zero magnetic field, we find that assuming a perfect Andreev reflection ($T^{he} = 1$) and no backscattering and cross Andreev reflection ($T^{car} = 0 = T^{b}$), the conductance is quantized: $G_{34} = 12e^2/h$, and agrees excellently with experimental observations. 

We now present simulations corresponding to finite voltage bias measurements, still at zero field. In this case, no Majorana Kramers pairs are expected to be present in the constriction.
Details about the simulation are presented in Appendix~\ref{appendix:bhz}.
Using $\phi = 0$, we show simulated $dI/dV$ (equivalent to $G_{34}$) as a function of $E$ in Fig.~\ref{fig:fig3}~(b) for different constriction widths $d$, comparing directly to the measured $dI/dV$ (brown line). When $d$ is large, a small mass gap in the constriction suppresses backscattering and leads to a strong Andreev response. As the constriction is narrowed, the Andreev reflection is weakened. This is qualitatively similar to the effect of a potential barrier on Andreev reflection in the BTK model of a normal metal-SC junction. Additionally, we find Andreev reflection associated with the superconducting constriction is robust to Anderson disorder. Figure~\ref{fig:fig3}~(c) shows $dI/dV$ in the presence of Anderson disorder (black line) with a potential amplitude $U_A = 60 \Delta$ averaged over 55 disorder realizations.
We use a random, uncorrelated on-site disorder potential sampled from a uniform distribution in the range $[-U_A/2, U_A/2].$
In Fig.~\ref{fig:fig3}~(c) we compare the experimental conductance at $V_g = -0.6$~V to the disorder-averaged conductance and find excellent quantitative agreement. 
We observe only slight deviations near the gap edge, most likely due to non-equilibrium effects and thermal broadening that are not captured 
in the Landauer-B\"{u}ttiker description.
Importantly, the nearly quantized three-terminal conductance we measured substantiates the existence of helical edge state transport in the device since trivial edge states or bulk carriers lack topological protection, generally leading to a non-quantized conductance that is sensitive to the voltage bias.

When a magnetic field is applied, a phase-bias will develop across the constriction and can lead to the formation of MKPs when $\phi = \pi$. In this case, perfect crossed Andreev reflection is expected at zero bias and results in a negative non-local $dI/dV$; although, at moderate field strengths, edge states will be gapped out leading to a suppression of $dI/dV$. From Fig.~\ref{fig:fig2}~(a), we see this clearly does not occur in our device.
A quantized Hall conductance anomalously robust under large magnetic fields has been reported in other InAs/GaSb double quantum wells for in-plane fields as large as 12 T~\cite{Du2015}.
This is in contrast with the general expectation that under a strong B field broken TRS should open a gap at the Dirac point of linearly dispersing edge states, eventually leading to the suppression of edge conductance.
While a topological excitonic insulating state in InAs/GaSb double quantum wells due to strong interlayer correlations may reconcile disagreement between theoretical predictions and experiment observations concerning temperature scaling of disorder effects~\cite{Pikulin2014, Du2017, Yu2018_exciton}, broken time-reversal symmetry is still expected to gap out the topological excitonic edge states.
A very likely scenario to resolve this discrepancy is one in which the Dirac point is moved down in energy (buried) due to higher-order corrections to the band structure of QSHIs~\cite{Li2018,Skolasinski2018}. We expect that this to be the case for our devices. Recent work theoretically investigated the impact of a buried Dirac point on Majorana zero modes~\cite{Schulz2020}, but study of MKPs in a similar vein is lacking. 
To assess the potential of a qubit based on MKPs, theoretical analysis of realistic models and experimental investigation of prospective devices are needed. 
MKPs have been predicted in QSHI-SC hetrostructures in which helical edge states hybridize at a QPC, see Fig.~\ref{fig:fig4}~(a). When the SC constriction is phase-biased by a flux $\Phi_0/2$ (where $\Phi_0$ is the superconducting magnetic flux quantum), MKPs can form at the ends of the constriction~\cite{Li2016, Pikulin2016}.
In the following, we will examine this case.

\begin{figure}[t]
    \centering
    \includegraphics[width=0.8\linewidth]{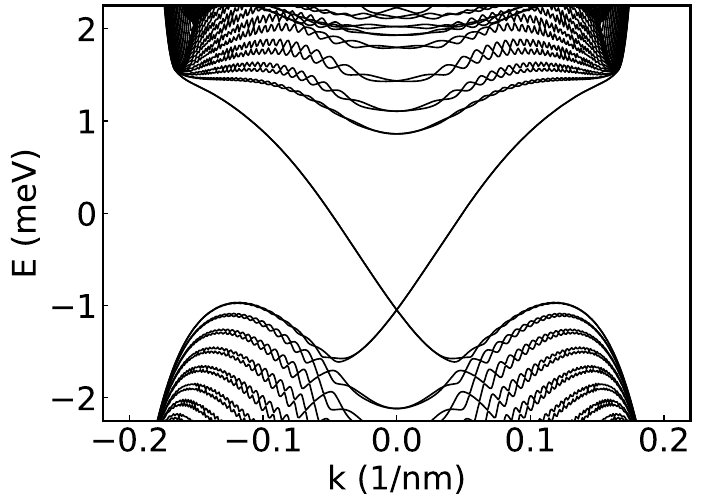} 
    \caption{\label{fig:bands}
    Dispersion of the L\"{o}wdin effective model for InAs/GaSb with a buried Dirac point.
    }
\end{figure}


\section{Majorana Kramers pairs with buried Dirac points}
\label{sec:mkp}
To investigate MKPs with and without a buried Dirac point, we consider a more detailed Hamiltonian for a similar inverted type-II InAs/GaSb bilayer derived from $\mathbf{k \cdot p}$ calculations using the L\"{o}wdin partitioning technique~\cite{Skolasinski2018}. This effective Hamiltonian is a variation of the BHZ model with additional spin-orbit coupling and $\sim \mathbf{k}^3$ momentum terms: $H_{\ell} = H_{bhz} + H_{soc} + H_{3}$ (see Appendix~\ref{appendix:lowdin} for details).
The Hamiltonian $H_{\ell}$ allows us to accurately model the two dimensional electron gas in the double quantum well without having to use two different Hamiltonians for the the InAs and GaSb layer and take into account the band-bending effects at the interface between the two layers~\cite{Liu1995}.
To compare the properties of MKPs with and without a buried Dirac point as directly as possible, we will consider two Hamiltonians:
$
H^{(1)} = H_{bhz} + 0.1 H_{soc}
$
and
$
H^{(2)} = H_{\ell},
$
where $H^{(1)}$ has an exposed Dirac point and $H^{(2)}$ has a buried one.
$H^{(1)}$ includes the term $0.1 H_{soc}$ in order to account for structural asymmetries~\cite{Liu2008} that give rise to a gap opening with the activation of a Zeeman field in the z-direction.
In our simulations, we use experimentally relevant parameters $d = 600$~nm, $L_{s,x} = 900$~nm, and a uniform spatial grid with resolution $a=3$~nm; for more details and the specific parameters of the L\"{o}wdin effective model, see Appendix~\ref{appendix:lowdin}. 

\begin{figure}[b]
    \centering
    \includegraphics[width=\linewidth]{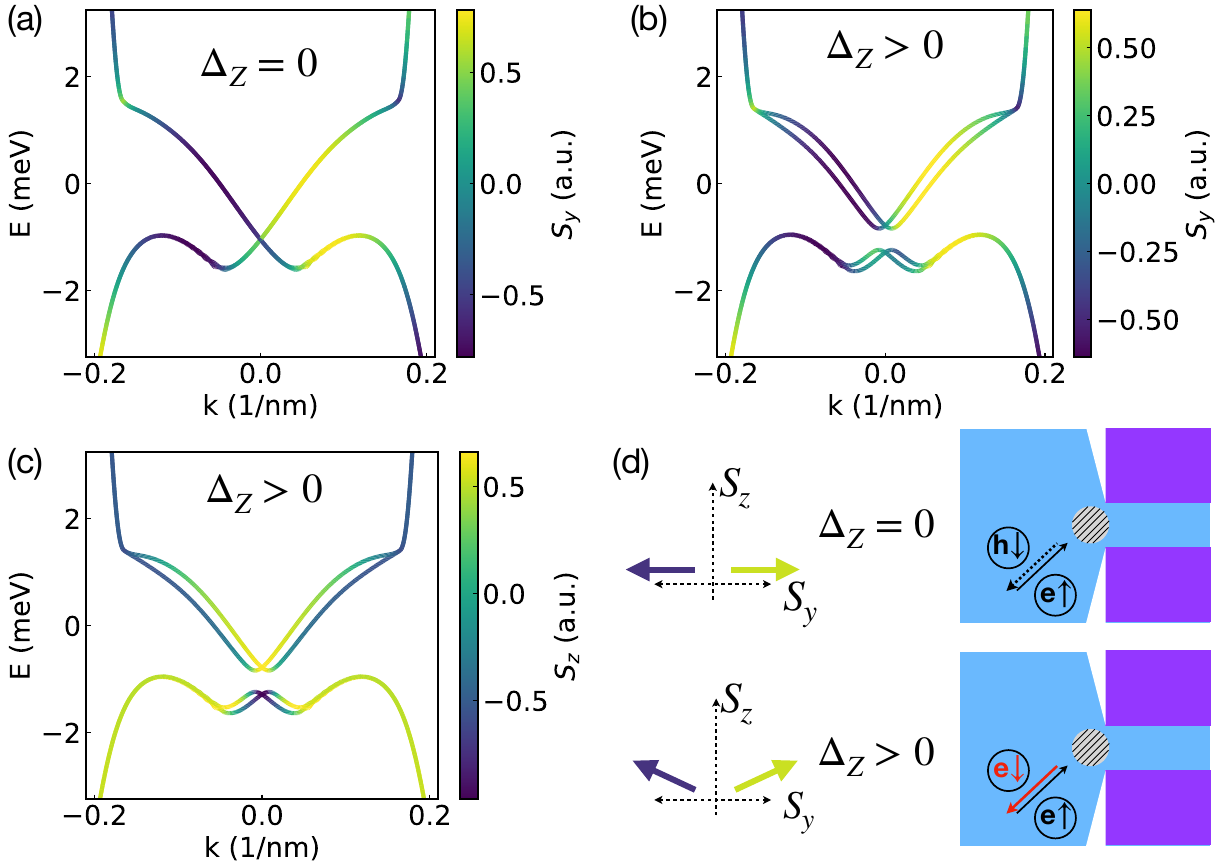} 
    \caption{\label{fig:spins}
    Edge state dispersion calculated from $H_{\ell}$ with spin projection $S_y$ for \textbf{(a)} $\Delta_Z = 0$ and
    \textbf{(b)} $\Delta_Z = 4\Delta$;
    \textbf{(c)} spin projection $S_z$ for $\Delta_Z = 4 \Delta$. 
    \textbf{(d)} Schematic of spin canting of edge modes with $\Delta_Z>0$ associated with an out-of-plane magnetic field leading to edge backscattering at the SC constriction.
    }
\end{figure}

Figure~\ref{fig:bands} shows the dispersion of a nanoribbon with a $1.2~\mu$m width calculated using $H_{\ell}$. We see that instead of the helical edge modes crossing in the middle of the bulk gap, the Dirac point lies buried below the bulk valence band. We verify the spin-momentum locking of the edge modes in Fig.~\ref{fig:spins}~(a) where edge states propagating with momentum $\pm k_x$ have spin projection $\pm S_y$. When an out-of-plane magnetic field is applied, a Zeeman splitting $\Delta_Z$ opens a minigap in the edge dispersion as shown in Fig.~\ref{fig:spins}~(b-c). Due to the magnetic field, we see that the edge states acquire a non-zero $S_z$ spin projection in Fig.~\ref{fig:spins}~(c) and the edge band degeneracy is lifted. In this case, the edge state spins begin to cant with respect to the $S_y$-axis, as depicted in Fig.~\ref{fig:spins}~(d), allowing for spin-flip scattering at the SC constriction.

\begin{figure}[b]
    \centering
    \includegraphics[width=\linewidth]{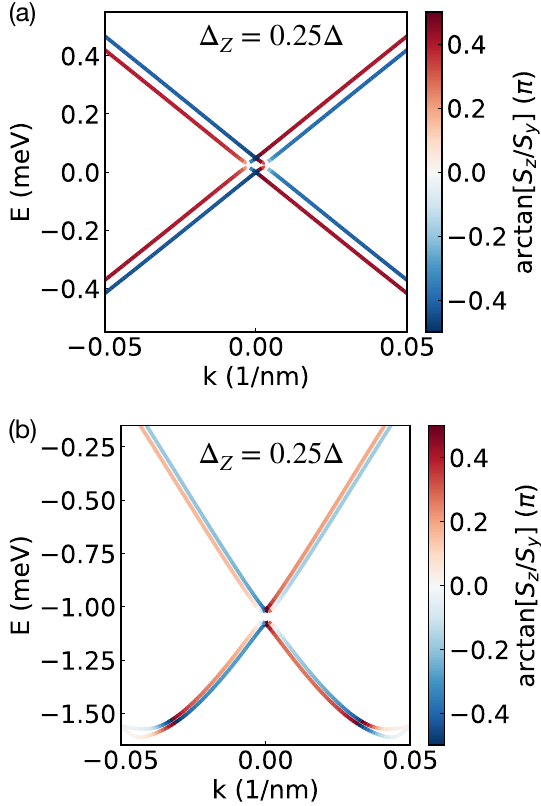} 
    \caption{\label{fig:spin_sus}
    Edge state dispersion with spin canting angle for $\Delta_Z = 0.25\Delta$ calculated from \textbf{(a)} $H^{(1)}$ (exposed Dirac point) and \textbf{(b)} $H^{(2)}$ (buried Dirac point).
    }
\end{figure}

We can analyze the effect of spin canting on edge conductance where
within the Born approximation we can estimate the edge state mean-free path to be~\cite{Pikulin2014}
$\ell_{mfp} \propto v_F/ \xi (\hbar \delta \mu/ V_0 \delta_B   )^2,
$
where
$
\langle V(r) V(r^{\prime}) \rangle = \xi V_0^2 \delta(r-r^{\prime})
$
describes uncorrelated disorder, $\delta \mu$ is the difference between the chemical potential and the position of the Dirac point, and $\delta_B$ is the minigap opened by the magnetic field. The constant of proportionality in this expression, in part, describes the spin susceptibility of the edge states i.e. how readily they align with the magnetic field. In Fig.~\ref{fig:spin_sus}, we quantify the spin susceptibility of $H^{(1)}$ and $H^{(2)}$ for $\Delta_Z = 0.25\Delta$ by comparing the relative spin angle between edge states with opposite momentum. We find that buried edge bands have stronger spin susceptibility where the spins of edge states with opposite momentum have a relative angle less than $\pi$, even in the middle of the bulk gap.
This suggests even weak disorder, such as the weak uncorrelated disorder we incorporated into our transport calculations, can cause non-negligible backscattering at low fields in our three-terminal measurement extending over roughly 8 microns.
This mechanism likely describes the slight dip in the sub-gap conductance observed in Fig.~\ref{fig:fig2}~(a) at low fields.

\begin{figure}[b]
    \centering
    \includegraphics[width=\linewidth]{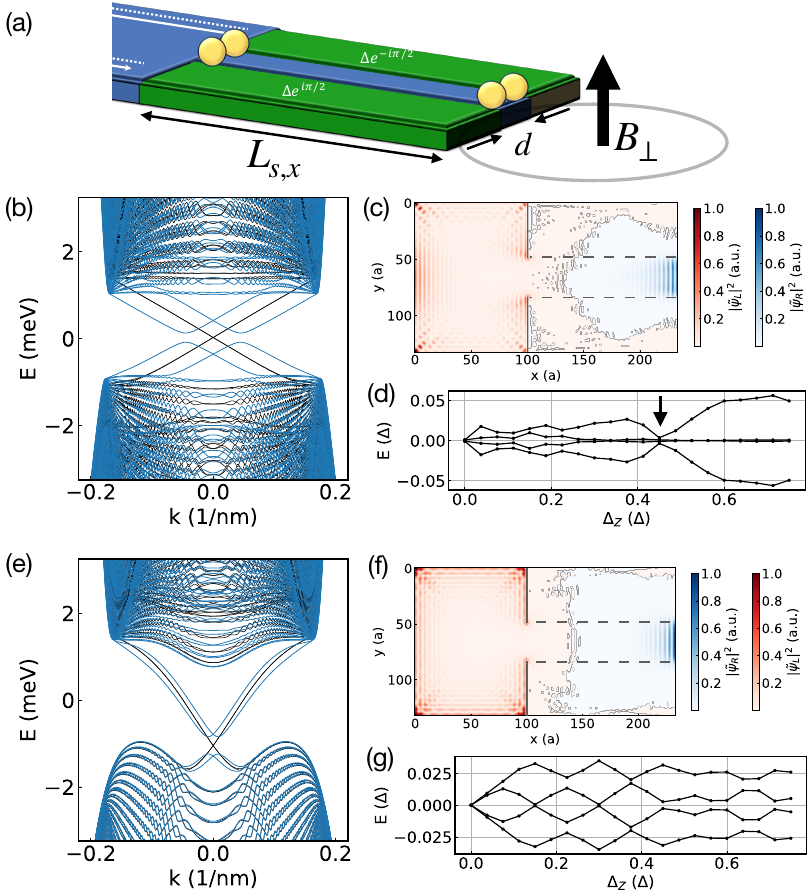} 
    \caption{\label{fig:fig4}
    \textbf{(a)} Schematic of the device setup where a $\pi$-phase difference across the junction leads to the creation of MKPs (yellow).
    \textbf{(b)} Dispersion of $H^{(1)}$ with (blue) and without (black) $\Delta_Z$.
    \textbf{(c)} Relative weights $\vert \tilde{\psi}_R\vert^2$ and $\vert \tilde{\psi}_L\vert^2$ of a MKP at $\Delta_Z = 0$ using $H^{(1)}$ and \textbf{(d)} corresponding eigenenergies versus $\Delta_Z$. The arrow indicates the onset of a spinless p-wave state.
    \textbf{(e)} Dispersion of the L\"{o}wdin effective model for InAs/GaSb with (blue) and without (black) $\Delta_Z$.
    \textbf{(f)} Relative weights $\vert \tilde{\psi}_R\vert^2$ and $\vert \tilde{\psi}_L\vert^2$ of a MKP at $\Delta_Z = 0$ using $H^{(2)}$ and \textbf{(g)} corresponding eigenenergies versus $\Delta_Z$. 
    }
\end{figure}

Figure~\ref{fig:fig4}~(b) presents the dispersion for $H^{(1)}$ with $\Delta_Z = 0$ and $4\Delta$. To visualize the MKPs, we consider the local density of states of a pair of MKP eigenstates transformed into left- and right-parity states $\psi_{L/R} = \frac{1}{\sqrt{2}} \left( \psi_+ \pm i \psi_- \right)$, where $\psi_{\pm}$ are one pair of positive and negative energy MKP eigenstates. This is a useful transformation of both Majorana zero modes and Kramers pairs since the left- and right-parity states are expected to be highly localized with little overlap between them. 
To be able to visualize using only positive effective weights the spatial profiles of the MKPs,
$\vert \psi_{L/R} \vert^2$ relative to their overlap: 
\begin{align}
    \vert \tilde{\psi}_{L/R} \vert^2 & = 
    \begin{cases}
    \pm \left( \vert \psi_L \vert^2 - \vert \psi_R \vert^2 \right), & \pm \left( \vert \psi_L \vert^2 - \vert \psi_R \vert^2 \right) \ge 0 \\
    0, & o.w.
    \end{cases}
\end{align}
Note that in the thermodynamic limit ($L_{s,x} \rightarrow \infty$), $\vert \tilde{\psi}_{L/R} \vert^2 = \vert \psi_{L/R} \vert^2$.
In Fig.~\ref{fig:fig4}~(c) we plot the MKP relative weights $\vert \tilde{\psi}_{L/R} \vert^2$ with $\Delta_Z = 0$. Compared to Fig.~\ref{fig:fig4}~(a), $\vert \tilde{\psi}_L\vert^2$ here is extended along the vacuum edge of the QSHI region. This occurs because the tunneling barrier (combination of edge state mass gap and constriction length separating the QPC and QSHI) between the MKP and fermionic modes in the QSHI region is insufficient to pin the left Majorana modes to the constriction.
The evolution of MKP energies with Zeeman splitting $\Delta_Z$ is presented in Fig.~\ref{fig:fig4}~(d).
With time-reversal symmetry lifted, the MKPs split from $E=0$ and modulate weakly before evolving near $\Delta_Z = 0.45\Delta$ into one pair of Majorana zero modes and a pair of trivial finite energy states.

Before analyzing the viability of MKPs with a buried Dirac point, let's discuss the impact of the constriction on edge states with an exposed Dirac point compared to a buried one.
When top and bottom edge states hybridize in a normal constriction, they open a gap $\Delta_{qpc}$ in the dispersion near the Dirac points. 
When the Dirac point is in the middle of the bulk gap, this can gap out the edge state dispersion if the constriction width is smaller than the penetration of the edge states into the bulk.
For edge states with a buried Dirac point, unless $\Delta_{qpc}$ exceeds the difference in energy between the Dirac point and the top of the valence band, edge states in the QPC remain gapless regardless of the position of the Fermi energy in the bulk gap.
Thus, the key difference arises in the role of the mass gap opening at the Dirac point when the edge states encounter the constriction.
Despite this modification to the bands, with a buried Dirac point there still exists two Fermi points with $k>0$ that have superconducting gaps of differing signs inside the $\pi$ phase-biased SC constriction. Then, in principle, pairing of counter-propagating edge states on the top and bottom of the constriction is unaffected by a buried Dirac point, permitting MKPs at the ends of the constriction.
Now, in the case of a buried Dirac point where the edge dispersion remains gapless in the constriction, no insulating domain wall exists between the QSHI and SC constriction that pins the MKP. 
Without a domain wall to pin the MKP, the tunneling between low energy fermionic modes in the QSHI and the MKP can be significant, leading to an \textit{extended} MKP state. This is similar to what happens to a Majorana zero mode at the end a topological superconductor (TSC) nanowire when it couples strongly to a normal region where the SC gap is removed (see Appendix~\ref{appendix:moredata}).
This can also occur with an exposed Dirac point (c.f. Fig.~\ref{fig:fig4}~(c)), but it will be ubiquitous with a buried Dirac point. Due to this strong coupling of the MKPs and fermionic states, a narrower constriction is needed in order to pin MKPs (see Fig.~\ref{fig:pinnedMKP}) and differentiate between conductance signatures of MKPs and trivial Andreev bound states.

Figure~\ref{fig:fig4}~(e) shows the dispersion of $H^{(2)}$ where the Dirac point lies just below the valence band edge for $\Delta_Z = 0$ and $\Delta_Z=4\Delta$. 
In Fig.~\ref{fig:fig4}~(f), we present the analog of panel (b) with $H^{(2)}$. 
The MKP weights are similar to the exposed Dirac point scenario in panel (c), but here the MKP is more localized along the vacuum edge of the QSHI region.
In Fig.~\ref{fig:fig4}~(g), we notice a qualitative difference in the effect of the Zeeman splitting: 
oscillations in the MKP energies are regular and persist throughout the range of Zeeman splittings we consider.
These oscillations are similar to the oscillations expected in finite-sized systems hosting Majorana zero modes which have been investigated extensively~\cite{Fleckenstein2018, Cao2019}, but here they correspond to MKPs rather than a single pair of Majorana zero modes.
Importantly, contrary to the case in panel (d) with an exposed Dirac point, in panel (g) no transition occurs to a spinless p-wave state (with a single pair of zero energy states) when the Dirac point is buried. 
In fact, the MKP weights at finite $\Delta_Z$ (see Appendix~\ref{appendix:moredata}) suggest the system still contains weakly hybridized MKPs despite lacking topological protection~\cite{Zhang2013}, even for values of $\Delta_Z$ where the $H^{(1)}$ system is driven into a spinless p-wave phase. 
This delay in transitioning to a spinless p-wave state can be traced back directly to the buried Dirac point.
For Majorana zero modes with buried Dirac points, it was shown that the transition from a trivial phase to a TSC induced by a Zeeman splitting is shifted to larger values of $\Delta_Z$ when the chemical potential does not coincide with the Dirac point~\cite{Schulz2020}. In our context, the transition from weakly hybridized MKPs to Majorana zero modes is shifted to larger $\Delta_Z$ when the chemical potential does not align with the Dirac point.
Thus, a buried Dirac point may help preserve the Majorana Kramers pairs under finite magnetic fields.

\section{Conclusions}
In this work, we present an experimental device specifically designed toward realizing MKPs and which satisfies the essential requirements toward this goal.
We fabricated a device based on a InAs/GaSb (15nm/5nm) bilayer with a superconducting Ta constriction. Using an electrostatic gate, we tuned the InAs/GaSb bilayer to a bulk band gap and measured a three-terminal conductance $dI/dV \approx 12(e^2/h)$ at zero field with a conductance plateau remaining robust up to an external magnetic field of 2~T.
Using a modified Landauer-B\"{u}ttiker analysis accounting for helical edge states and Andreev processes, we found the measured $dI/dV$ consistent with $\sim 98 \%$ Andreev retroreflection probability which is in quantitative agreement with tight binding simulations of the device.
The robustness of the conductance to B strongly points to a buried Dirac point.

Turning to a theoretical analysis of potential future devices, our simulations show a buried Dirac point in our device's configuration works to \textit{preserve} MKPs under broken TRS. 
We have shown that a buried Dirac point also reduces the tunneling barrier between the QSHI and the SC constriction, which can cause
any resonant structure associated with the MKPs, such as crossed Andreev reflection~\cite{Li2016, Pikulin2016},
to be washed out. This low tunneling barrier consequentially unpins a MKP (the left side of the constriction in Fig.~\ref{fig:fig4}~(c,f)), resulting in a non-local MKP that remains decoupled from the other MKP. Since the coupling of MKPs on either end of the constriction is unaffected, the topological gap is unaffected. 
Given that, the MKP can become an extended edge state rather than a localized ``defect" and we can apply edge state techniques such as scattering at normal QPCs to probe the MKPs. On the other hand, if we want to pin the MKP as originally proposed, this can be achieved by reducing the separation of the Ta electrodes further to less than 100~nm, well within the resolution of available lithographic techniques. Future work on devices with a narrower Ta constrictions may be promising for finding evidence for perfect Andreev reflection resonant with MKPs in non-local conductance measurements. 

\section*{Acknowledgements} 
J.J.C. acknowledges helpful discussions with Rafal Skolasinski. J.C.C. is supported by a LDRD project.
Work by E.R. and W.P. was funded by the US Department of Energy, Office of Basic Energy Sciences, via Award DE-SC0022245 (theoretical modeling, device fabrication, measurements, data analysis).
Materials growth was supported by the LDRD program at Sandia. Device fabrication was carried out at CINT, a user facility by DOE.
Sandia National Laboratories is a multi-mission laboratory managed and operated by National Technology \& Engineering Solutions of Sandia, LLC (NTESS), a wholly owned subsidiary of Honeywell International Inc., for the U.S. Department of Energy’s National Nuclear Security Administration (DOE/NNSA) under contract DE-NA0003525. This written work is authored by an employee of NTESS. The employee, not NTESS, owns the right, title and interest in and to the written work and is responsible for its contents. Any subjective views or opinions that might be expressed in the written work do not necessarily represent the views of the U.S. Government. The publisher acknowledges that the U.S. Government retains a non-exclusive, paid-up, irrevocable, world-wide license to publish or reproduce the published form of this written work or allow others to do so, for U.S. Government purposes. The DOE will provide public access to results of federally sponsored research in accordance with the DOE Public Access Plan.
\\

\appendix

\section{Additional Experimental Data}\label{appendix:exp}
\begin{figure}[h!!!!]
    \centering
    \includegraphics[width=0.6\linewidth]{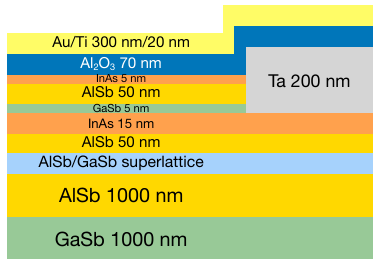} 
    \caption{\label{fig:stack}
    Heterostructure material stack of the superconducting device.
    }
\end{figure}

Figure~\ref{fig:stack} shows the full material stack of the superconducting device discussed in the main text.
A HAADF STEM image of a heterostructure we grew using the same growth parameters (e.g., the same growth temperatures) and same MBE machine as the one we used in our manuscript shows perfect epitaxial interfaces between InAs, GaSb, and AlSb, respectively~\cite{Pan2025}. This clearly demonstrates the high quality of the interfaces in our InAs/GaSb samples. Because of this we conclude that in for our devices the dominant disorder arises from random, screened, charge impurities and other sources of disorder, such as interfacial disorder~\cite{Magri2002, Riney2025}, should be negligible.

\begin{figure}[h!!!!]
    \centering
    \includegraphics[width=0.98\linewidth]{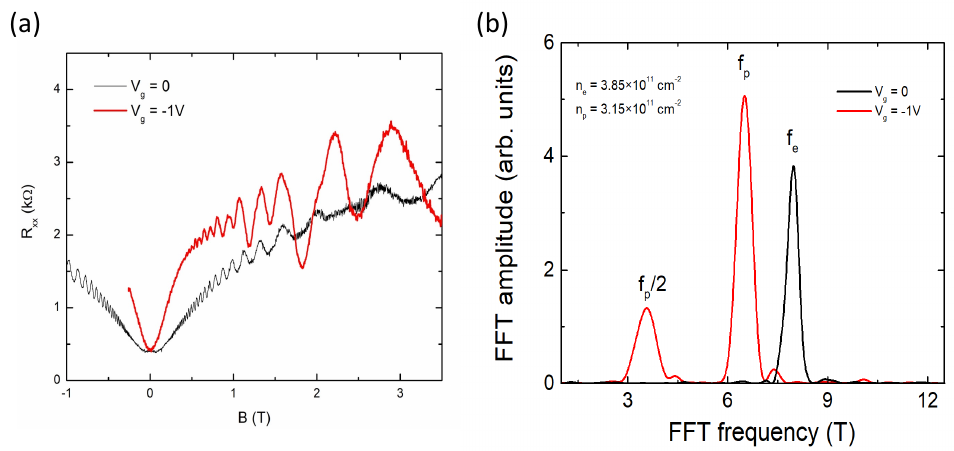} 
    \caption{\label{fig:magnetoR}
    Electronic transport characterization in the device region without the superconductor. (a) Magnetoresistance $R_{xx}$ at two gate bias voltages of 0 (black curve) and -1 V (red curve), respectively. Shubnikov-de Haas (SdH) oscillations are observed. (b) Fast Fourier Transform (FFT) analysis of SdH oscillations in panel (a). At V$_g$ = 0 V, where the carriers are of electron type, a single FFT peak is seen. At V$_g$ = -1 V, where the carriers are of hole type, two peaks are seen. From the positions of the main peaks, the electron and hole densities are deduced: $n_e$ = 3.85$\times$10$^{11}$ cm$^{-2}$ and $n_p$ = 3.15$\times$10$^{11}$ cm$^{-2}$. 
    }
\end{figure}

Magnetoresistance ($R_{xx}$) data in the device region without the superconductor is shown in Fig.~\ref{fig:magnetoR}. At both $V_g$ = 0 and -1 V, the Shubnikov-de Haas oscillations are observed (Fig.~\ref{fig:magnetoR}a). Fast Fourier Transform (FFT)  analysis of these SdH oscillations is carried out, from which the electron density $n_e$ = 3.85$\times$10$^{11}$ cm$^{-2}$ at $V_g$ = 0V and hole density $n_p$ = 3.15$\times$10$^{11}$ cm$^{-2}$ at $V_g$ = -1V are deduced. Furthermore, their mobilities are obtained. At $V_g$ = 0 V, $\mu_e$ = 8.1×10$^4$ cm$^{2}$/Vs; at $V_g$ = -1 V, $\mu_p$ = 9.4$\times$10$^4$ cm$^{2}$/Vs. The calculated electron and hole mean free path is 0.83 $\mu$m and 0.87 $\mu$m, respectively. These mean free paths are smaller than the device dimension ($\sim$3 $\mu$m). So, the bulk contribution to $R_{xx}$ can be non-negligible.

\begin{figure}[h!!!!]
    \centering
    \includegraphics[width=0.6\linewidth]{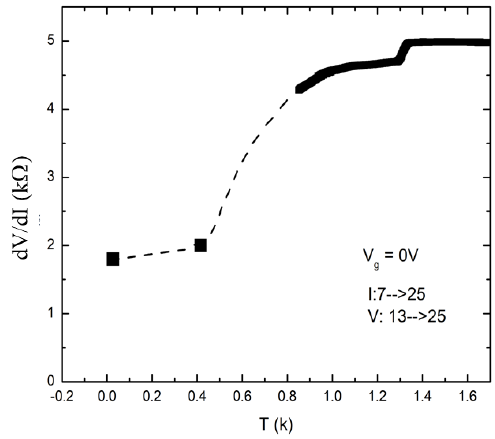} 
    \caption{\label{fig:SCdata}
    Temperature dependence of the three-terminal differential resistance dV/dI at $V_g$ = 0 V. The dashed line is a guide to the eye. 
    }
\end{figure}

In Fig.~\ref{fig:SCdata}, we present three-terminal differential resistance dV/dI at $V_g$ = 0 V measured as a function of sample temperature (T). It is constant at high temperatures. The sharp drop at T $\sim$ 1.3 K is due to the onset of the superconducting transition of the Ta electrode. dV/dI continues decreasing as the temperature is lowered.

\begin{figure}[h]
    \centering
    \includegraphics[width=0.6\linewidth]{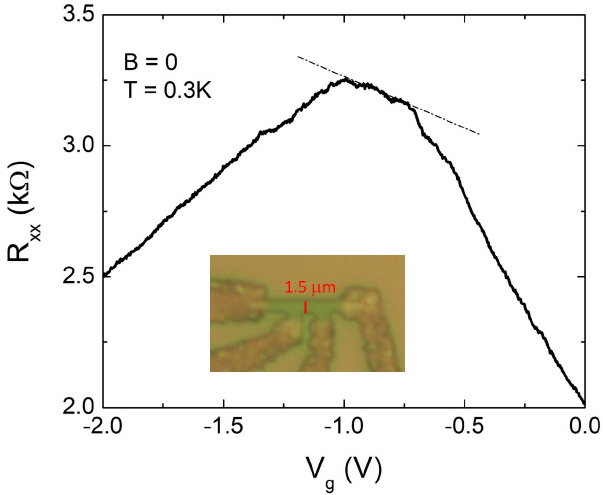} 
    \caption{\label{fig:pibar}
    $R_{xx}$ measured in $\pi$-bar device (inset).
    }
\end{figure}

In Fig.~\ref{fig:pibar}, we present $R_{xx}$ measurements as a function of gate voltage in a $\pi$-bar device (inset). The device was fabricated on a different sample than the one presented in the main text, but both samples follow the same fabrication procedure. We observe a peak $R_{xx} \approx 3.5$~k$\Omega$, which is consistent with $R_{xx}$ measured in other InAs/GaSb double quantum wells in the QSHI regime with similar device geometry~\cite{Knez2011}.

In Fig.~\ref{fig:network_model}, we show the resistor network model used to estimate the expected $R_{xx}$ in the QSHI phase. The edge modes between contacts are assumed to have resistance $R_0$ with ideal contacts, which results in a resistance across contacts 3 and 13: $R_{xx} = \frac{V_{3-13}}{I} = \frac{3}{8}R_0 \approx 4.84~k\Omega$. There are two possibilities we've identified to explain the discrepancy between the expected and measured $R_{xx}$. First, there may be a parallel bulk channel with resistant $R_{bulk} \sim 2.74~k\Omega$. Alternatively, the assumption of an ideal contact may break down for lead 25 where Ta forms a constriction, in which case a contact resistance of $\sim 13 R_0$ can account for the discrepancy. Both effects likely contribute to the observed $R_{xx}$, but the nearly quantized value of $dI/dV$ ($12e^2/h$) in the three-terminal measurement is consistent with a primarily edge state picture.

\begin{figure}[h!!!!]
    \centering
    \includegraphics[width=0.8\linewidth]{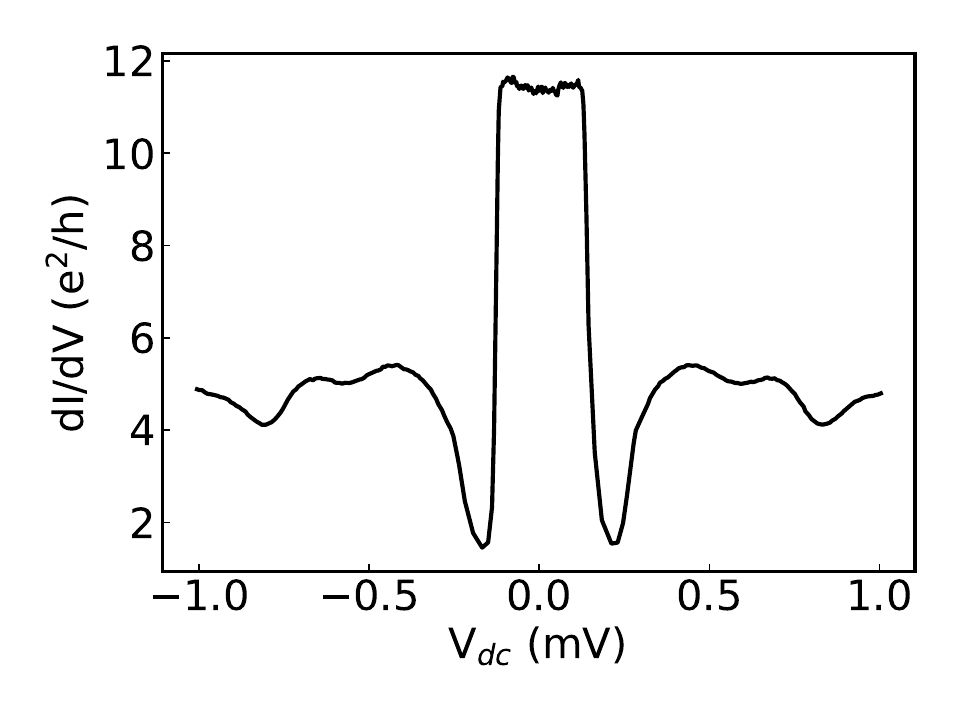} 
    \caption{\label{fig:secondcooldown}
    Three-terminal $dI/dV$ vs voltage bias between contacts 13 and 25 at $B=0$ and the gate set to the charge neutrality point upon a second cooldown of the device.
    }
\end{figure}

\begin{figure}[h!!!!]
    \centering
    \includegraphics[width=0.98\linewidth]{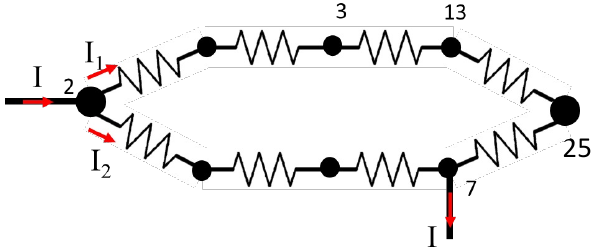} 
    \caption{\label{fig:network_model}
    Resistor network model for edge transport of the device. Contacts are indicated by dots and the contact labeling is the same as shown in Fig.~\ref{fig:fig1}(c).
    }
\end{figure}

Lastly, we performed a second  cooldown of the device reported in the main text to check reproducibility of the zero field sub-gap conductance plateau. The non-local differential conductance measurement is shown in Fig.~{\ref{fig:secondcooldown}}.

\section{BHZ Tight Binding Model}\label{appendix:bhz}

\begin{figure}[h]
    \centering
    \includegraphics[width=0.8\linewidth]{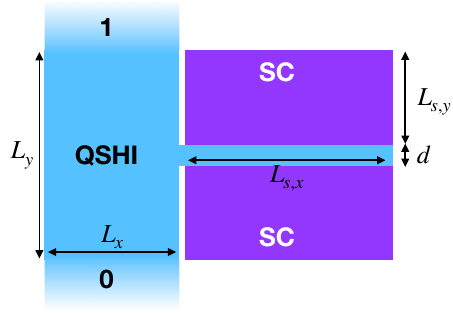} 
    \caption{\label{fig:dimensions}
    Schematic of system geometry used in simulations.
    }
\end{figure}

To simulate the conductance across the QPC, we use the following tight binding Bogoliubov-de Gennes Hamiltonian on a square lattice:
\begin{widetext}
\begin{align}
   H^{(BdG)}_{bhz}  =& \sum_{\mathbf{r}_n} \psi^{\dagger}_{\mathbf{r}_n} \left( \mu_{sm} \tau_z \otimes \sigma_0 \otimes \rho_z - \mu \tau_z \otimes \sigma_0 \otimes \rho_0 - \Delta(\mathbf{r}_n)\tau_y \otimes \sigma_y \otimes \rho_0  \right) \psi_{\mathbf{r}_n} \nonumber \\
   &- \sum_{\left< nm \right>} \delta_{y_n,y_m} \psi^{\dagger}_{\mathbf{r}_n} \left( \frac{t_1}{2}\tau_z \otimes \sigma_0 \otimes \rho_z + \frac{t_2}{2} \tau_z \otimes \sigma_0 \otimes \rho_0 + \frac{i\lambda}{2} \tau_0 \otimes \sigma_z \otimes \rho_x \right)
   \psi_{\mathbf{r}_m} \nonumber \\
   &- \sum_{\left< nm \right>} \delta_{x_n,x_m} \psi^{\dagger}_{\mathbf{r}_n} \left( \frac{t_1}{2}\tau_z \otimes \sigma_0 \otimes \rho_z + \frac{t_2}{2} \tau_z \otimes \sigma_0 \otimes \rho_0 - \frac{i\lambda}{2} \tau_z \otimes \sigma_0 \otimes \rho_y \right)
   \psi_{\mathbf{r}_m},
   \label{eq:BHZ}
\end{align}
where 
$
\psi_{\mathbf{r}_n} = (c_{\mathbf{r}_n},~c^{\dagger}_{\mathbf{r}_n})^T,
$
$
c_{\mathbf{r}_n} = (c_{\mathbf{r}_n,E_1\uparrow},~c_{\mathbf{r}_n,H_1\uparrow},~c_{\mathbf{r}_n,E_1\downarrow},~c_{\mathbf{r}_n,H_1\downarrow})^T,
$
$c^{\dagger}_{\mathbf{r}_n,\rho \sigma}$ ($c_{\mathbf{r}_n,\rho \sigma}$) is the creation (annihilation) operator for an electron at site $\mathbf{r}_n$ in orbital $\rho$ with spin $\sigma$, and
$\tau_i$, $\rho_i$ and $\sigma_i$ are $2\times 2$ Pauli matrices.
$\mu$  is the chemical potential, assumed to be uniform throughout the system. The superconducting gap $\Delta({\bf r})$ is taken to be $\Delta e^{i\phi/2}$ in the top superconductor, $\Delta e^{-i \phi/2}$ in the bottom superconductor, and zero otherwise.
Due to the bulk and structural inversion asymmetries in the double quantum well, we include the term
\begin{align}
   & H^{(BdG)}_{IA}  = \sum_{\mathbf{r}_n} \psi^{\dagger}_{\mathbf{r}_n} \left( - \Delta_{BIA}\tau_z \otimes \sigma_y \otimes \rho_y  \right) \psi_{\mathbf{r}_n} \nonumber \\
   &+ \frac{i}{2}\sum_{\left< nm \right>} \delta_{y_n,y_m} \psi^{\dagger}_{\mathbf{r}_n} \left( \mathrm{Re} (\chi) \tau_z \otimes \sigma_y \otimes \rho_+ + \mathrm{Im} (\chi) \tau_0 \otimes \sigma_x \otimes \rho_+ -  \tau_0 \otimes \sigma_x \otimes (t_e\rho_+ + t_h \rho_-) \right)
   \psi_{\mathbf{r}_m} \nonumber \\
   &- \frac{i}{2}\sum_{\left< nm \right>} \delta_{x_n,x_m} \psi^{\dagger}_{\mathbf{r}_n} \left( \mathrm{Re} (\chi) \tau_0 \otimes \sigma_x \otimes \rho_+ - \mathrm{Im}(\chi) \tau_z \otimes \sigma_y \otimes \rho_+ -  \tau_z \otimes \sigma_y \otimes (t_e\rho_+ - t_h \rho_-) \right)
   \psi_{\mathbf{r}_m} ,
\end{align}
\end{widetext}
where $\Delta_{BIA}$, $t_e$ and $t_h$ describe the bulk inversion asymmetry and $\chi$ describes the structural inversion asymmetry. The dimensions of the system modeled are shown in Fig.~\ref{fig:dimensions}.
\begin{table*}[t]
\centering
\begin{tabular}{||c | c | c | c | c | c | c | c | c | c | c | c | c ||} 
 \hline
 $\mu_{sm}$ & $\Delta$ & $t_1$ & $t_2$ & $\lambda$ & $\Delta_{BIA}$ & $\chi$ & $t_e$ & $t_h$ & $L_x$ & $L_y$ & $L_{s,x}$ & $L_{s,y}$ \\ [0.5ex] 
 \hline
 0.5 & 0.05 & 1 & 0.6 & 0.3 & 0.01 & 0.06 & 0.003 & 0.003 & 80 & 140 & 100 & 40 \\ 
 \hline
\end{tabular}
\caption{Parameters used to generate Fig.~3 in the main text.}
\end{table*}

\section{L\"{o}wdin effective Tight Binding Model}\label{appendix:lowdin}
We model the buried Dirac point in InAs/GaSb using the effective Hamiltonian derived in Ref.~\cite{Skolasinski2018}:
\begin{align}
   H^{(BdG)}_{\ell}  =& \frac{1}{2} \sum_{\bf k} \Psi_{\bf k}^{\dagger} \left( h_{bhz} + h_{3} + h_{soc} + h_{\Delta} \right) \Psi_{\bf k},
\end{align}
where, using abbreviated notation for Kronecker products, we have
\begin{widetext}
\begin{align}
    h_{bhz} & = \tau_z \left( C - \mu + M\rho_z - Bk^2 \rho_z - Dk^2 + A(k_x \sigma_z \rho_x - k_y \rho_y) \right) + \frac{\Delta_Z}{2} \sigma_z \\
    h_3 & = \tau_z \left( F(k_x^3\sigma_z \rho_x - k_y^3 \rho_y) + Q (k_x^2k_y \rho_y - k_x k_y^2 \sigma_z \rho_x) \right) \\
    h_{soc} & = \tau_z \left( T(k_x \sigma_y  - k_y \sigma_z )\rho_+ 
                - G (k_x^3\sigma_y - k_y^3\sigma_x )\rho_+ 
                - W (k_xk_y^2 \sigma_y - k_x^2 k_y \sigma_x) \rho_+
                - H (k_x^3 \sigma_y + k_y^3 \sigma_x)\rho_- \right)
                \nonumber \\
                & + \tau_z \left( R (k_xk_y^2 \sigma_y + k_x^2 k_y \sigma_x) \rho_-
                + E_1 (k_x^2 - k_y^2)\sigma_x \rho_y + E_2 k_x k_y \sigma_y \rho_y \right) \\
    h_{\Delta} & = \begin{pmatrix}
    0 & \Delta \\
    \Delta^* & 0
    \end{pmatrix} \sigma_0 \rho_0,
\end{align}
\end{widetext}
and 
$
\Psi_{\bf k} = (c_{\bf k }, (-i\sigma_y)c_{\bf k}^{\dagger})
$
with 
$
c_{\bf k} = (c_{\mathbf{k},E_1\uparrow},~c_{\mathbf{k},H_1\uparrow},~c_{\mathbf{r},E_1\downarrow},~c_{\mathbf{k},H_1\downarrow})^T
$. The matrices $\tau_i,~\sigma_i$~and $\rho_i$ are Pauli matrices in Nambu, spin, and orbital spaces, respectively, with $\rho_+ = \mathrm{diag}\{1,~ 0\}$ and $\rho_- = \mathrm{diag}\{0,~ 1\}$. 
Similar to the approach to the conductance calculations, we Fourier transform Hamiltonian into $\bf r$-space to perform tight binding calculations.
Parameters used in calculations are shown in Tables I. and II. The parameters used for simulations with $H^{(BdG)}_{\ell}$ correspond to an InAs/GaSb heterostructure with layer thicknesses 12.5 nm and 5 nm, respectively. These thicknesses differ from those in our experiment (15/5 nm) which suggests there will be differences in the bandstructure between the two heterostructures. Given other effects such as strain from AlSb layers with different thicknesses are not completely accounted for, we believe the differences between the two model are minor if (i) the magnitudes of the bulk gaps are consistent ($\sim 3$~meV) and (ii) they both reside in the ``deeply inverted" regime where the Dirac point is buried.

\begin{table*}[t]
\centering
\begin{tabular}{|| c || c | c | c | c | c | c | c | c | c | c | c | c | c | c | c | c | c ||} 
 \hline
   Hamiltonian & $\mu$ & $\Delta$ & $A$ & $B$ & $C$ & $D$ & $E_1$ & $E_2$ & $F$ & $G$ & $H$ & $M$ & $Q$ & $R$ & $T$ & $W$ & $a$ \\ [0.5ex] 
\hline
 $H^{(1)}$(12.5/5 nm) & 108 & 0.8 & -6.2 & -273.4 & 98.4 & -116.5 & 3.4 & 10.1 & 0 & 6.5 & -8.95 & -18.5 & 0 & -35.8 & -0.16 & 6.73 & 3.0 \\ 
 \hline
 $H^{(2)}$(12.5/5 nm) & 108 & 0.8 & -6.2 & -273.4 & 98.4 & -116.5 & 32.8 & 96.4 & -170.2 & 62.5 & -85.5 & -18.5 & 179.4 & -341.9 & -1.6 & 64.3 & 3.0 \\ 
 \hline
\end{tabular}
\caption{Parameters used to generate Fig.~4 in the main text. Energies in units of eV and lengths in units of nm.}
\end{table*}

\section{Conductance calculations}\label{appendix:transport}
In the four-terminal setup presented in the main text, we assume contact 4 is a grounded superconductor. Then we can generalize the LB method to account for Andreev processes~\cite{Hatefipour2022}:
\begin{align}
    I_i & = \sum_{j=1}^N a_{ij} \left(V_j - V \right),
\end{align}
where $V$ is the voltage of the superconducting lead and
\begin{align}
    a_{ij} & = \frac{e^2}{h} \left( N_i \delta_{ij} - T^{ee}_{ij} + T^{he}_{ij} \right),
\end{align}
where $T^{ee}_{ij}$ is the electron transmission from lead $j$ to lead $i$, $T^{he}_{ij}$ is the transmission of an electron from lead $j$ to a hole in lead $i$ (i.e. Andreev reflection), and 
\begin{align}
    N_{i} & = \sum_{j=1}^{m} \left( T^{ee}_{ij} + T^{he}_{ij} \right),
\end{align}
is the total number of modes in lead $i$. 

We assume the energy of electrons injected from normal contacts is below the superconducting gap so that the current in lead 4 is entirely due to Andreev reflection. Furthermore, we assume helical edge states are the only allowed modes in the system i.e. ignore bulk contributions. 

Time-reversal symmetry gives us
\begin{align}
    T^{ee}_{33} & = T^{ee}_{11} := T^b, \qquad T^{he}_{33} = T^{he}_{11} := T^{he}, \\
    T^{ee}_{13} & = T^{ee}_{31} := T^{co}, \qquad T^{he}_{13} = T^{he}_{31} := T^{car},
\end{align}
where $0 \le T^{b},T^{he},T^{car},T^{co} \le 1$, and quasiparticle number conservation gives us
\begin{align}
    T^{b} + T^{co} + T^{he} + T^{car} & = 1.
\end{align}
Then we can derive the conductance matrix for the Hall bar by expressing each current in terms of the voltages of the normal contacts. For contact 1
\begin{widetext}
\begin{align}
    I_1 & = a_{11} V_1 + a_{12} V_2 + a_{13} V_3 - (a_{11} + a_{12} + a_{13}) V \\
    & = \frac{e^2}{h} \left[ (2 - T^{ee}_{11} - T^{he}_{11}) V_1 - ( T^{he}_{12} - T^{ee}_{12} ) V_2 + ( T^{he}_{13} - T^{ee}_{13} ) V_3 - ( 2 - T^{ee}_{11} - T^{he}_{11} + T^{he}_{12} - T^{ee}_{12} + T^{he}_{13} - T^{ee}_{13} ) V \right] \\
    & = \frac{e^2}{h} \left[ (2 + T^{he} - T^b) V_1 - V_2 + (T^{car} - T^{co})V_3 - 2(T^{car} + T^{he}) V \right].
\end{align} 
Similarly
\begin{align}
    I_2 & = \frac{e^2}{h} \left[ 2V_2 - V_1 - V_3 \right] \\
    I_3 & = \frac{e^2}{h} \left[ (2 + T^{he} - T^b) V_3 - V_2 + (T^{car} - T^{co})V_1 - 2(T^{car} + T^{he}) V \right].
\end{align}

Then the conductance can be found by solving the matrix equation
\begin{align}
    \begin{pmatrix}
        I_1 + 2T^A V \\ I_2 \\ I_3 + 2T^A V
    \end{pmatrix}
     & = \begin{pmatrix}
         2 + T^{he} - T^b & -1 & T^{car} - T^{co} \\
         -1 & 2 & -1 \\
         T^{car} - T^{co} & -1 & 2 + T^{he} - T^b
     \end{pmatrix}
     \begin{pmatrix}
         V_1 \\ V_2 \\ V_3
     \end{pmatrix},
\end{align}
\end{widetext}
where 
$
T^A = T^{he} + T^{car}.
$

\section{Additional Simulation Data}\label{appendix:moredata}
\begin{figure}[h!!!!]
    \centering
    \includegraphics[width=0.8\linewidth]{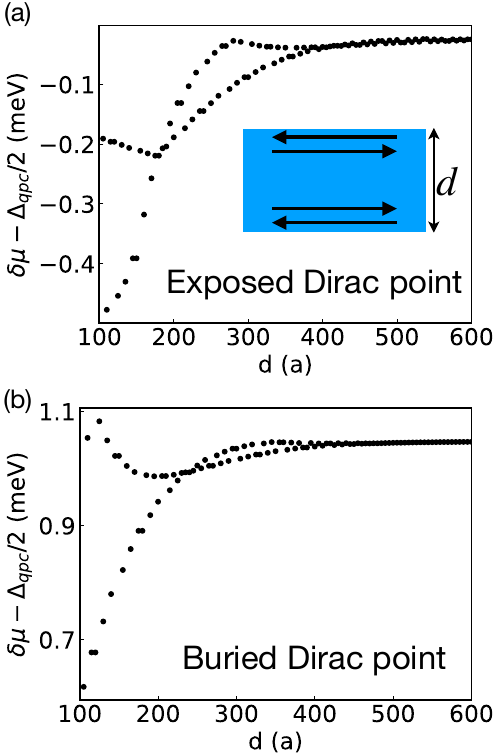} 
    \caption{\label{fig:back_gap} The difference between $\mu$ and the minigap edge opened at the Dirac point due to backscattering in a narrow channel of width $d$ using \textbf{(a)} $H^{(1)}$ and \textbf{(b)} $H^{(2)}$.
    }
\end{figure}

We can quantify the tunneling barrier between the MKP and vacuum edge states by calculating the difference between the chemical potential and the mini-gap edge in the edge dispersion: $\delta \mu - \frac{\Delta_{qpc}}{2}$. When $\delta \mu - \frac{\Delta_{qpc}}{2} < 0$, the edge states are gapped out at the chemical potential and a corresponding tunneling barrier exists. If $\delta \mu - \frac{\Delta_{qpc}}{2}>0$, then edge states exist at the chemical potential and nominally there is no tunneling barrier. This difference for exposed and buried Dirac points is shown in Fig.~\ref{fig:back_gap} as a function of the width of the nanoribbon $d$.

\begin{figure}[h!!!!]
    \centering
    \includegraphics[width=0.98\linewidth]{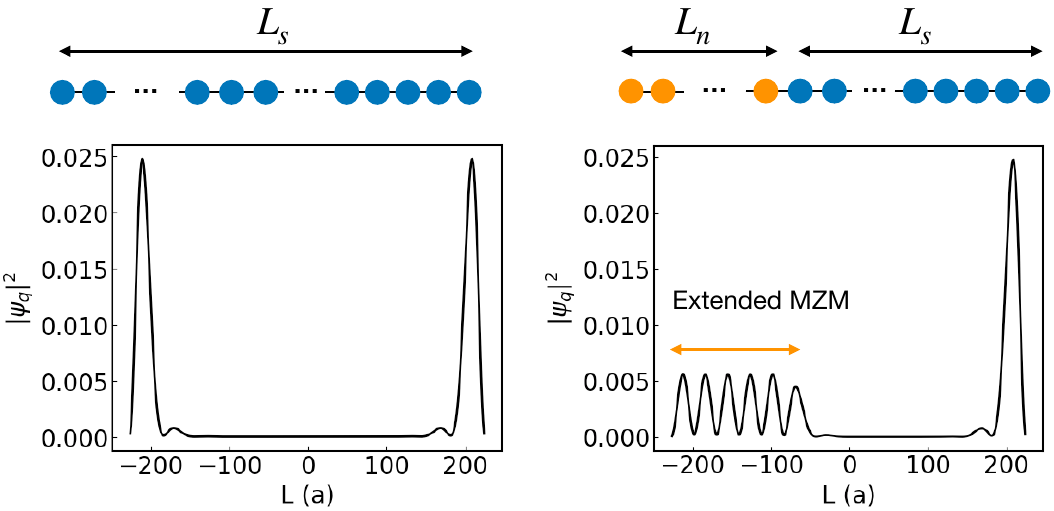} 
    \caption{Illustration of extended Majorana zero modes using the Lutchyn-Oreg nanowire model. In the right figure, the superconducting gap is suppressed in the range $-250a \le x \le -50 a$. In the absence of a potential barrier at the NS interface, the left MZM hybridizes with fermionic modes in the normal metal region leading to an extension of the left MZM.
    }
\end{figure}

\begin{figure}[h!!!!]
    \centering
    \includegraphics[width=0.98\linewidth]{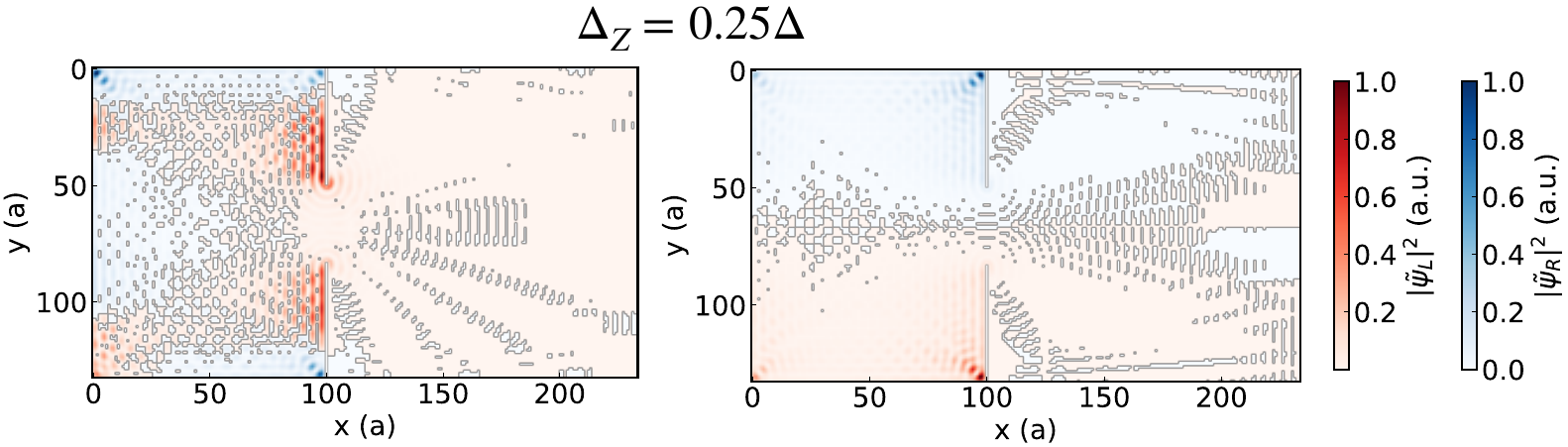} 
    \caption{Evolution of MKP wave function weights under Zeeman splitting for $H^{(1)}$ (exposed Dirac point).
    }
\end{figure}

\begin{figure}[h!!!!]
    \centering
    \includegraphics[width=0.98\linewidth]{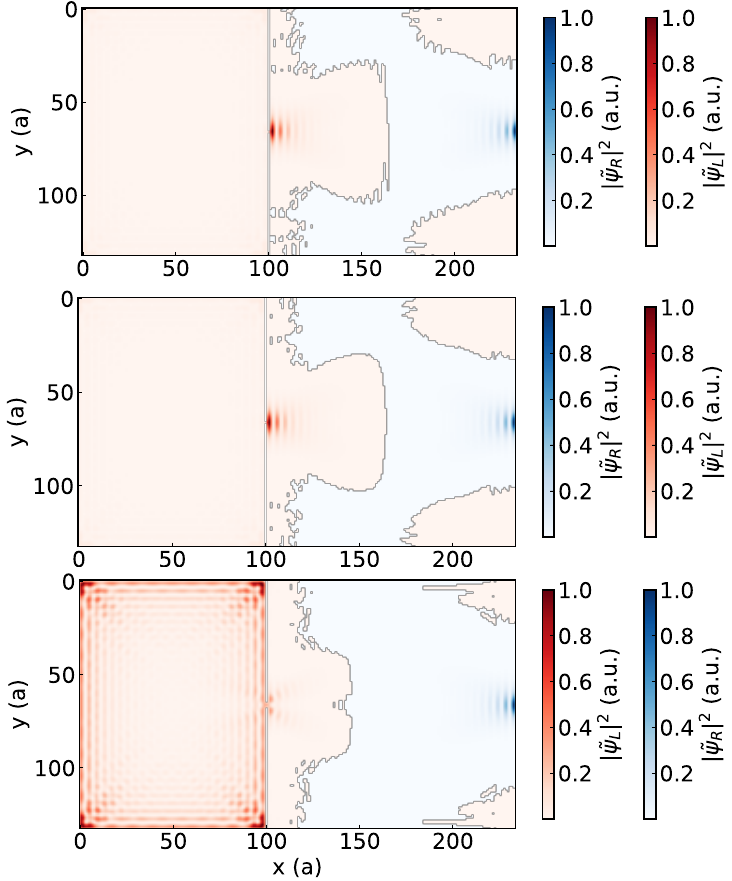} 
    \caption{\label{fig:pinnedMKP} Evolution of pinned to unpinned MKPs with increasing $d$ (top to bottom) for $H^{(2)}$ (buried Dirac point).
    }
\end{figure}

\begin{figure}[h!!!!]
    \centering
    \includegraphics[width=0.98\linewidth]{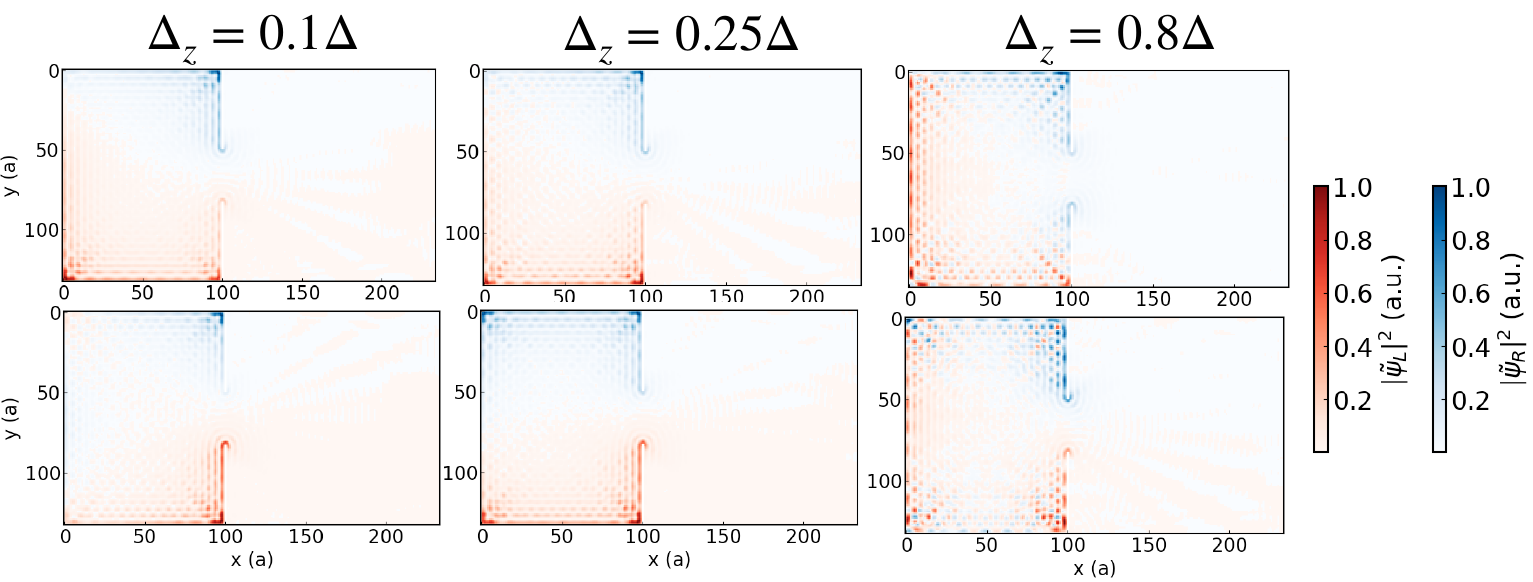} 
    \caption{Evolution of MKPs with increasing $\Delta_z$ for $H^{(2)}$ (buried Dirac point).
    }
\end{figure}

\bibliography{refs_Cuozzo}

\end{document}